\documentclass[draft,12pt,a4paper]{article}

\usepackage{fullpage}
\usepackage{latexsym}
\usepackage{amsfonts}
\usepackage{color}    
\usepackage{verbatim} 

\newcommand{\ket}[1]{\ensuremath{\left|#1\right>}}

\title{\bf Improved Bounds for the Approximate QFT}
\author{Donny Cheung\\[1ex]
\normalsize Department of Combinatorics and Optimization\\
\normalsize University of Waterloo\\
\normalsize Waterloo, ON, Canada N2L 3G1\\
{\normalsize\it e-mail: \tt dccheung@uwaterloo.ca}}
\date{}

\begin{document}

\maketitle

\begin{abstract}

It has previously been established that the logarithmic-depth 
approximate quantum Fourier transform (AQFT) provides a suitable replacement
for the regular QFT in many quantum algorithms.  Since the AQFT is less
accurate by definition, polynomially many more applications of the AQFT are
required to achieve the original accuracy.  However, in many quantum
algorithms, the smaller size of the AQFT circuit yields a net improvement over
using the QFT.

This paper presents a more thorough analysis of the AQFT circuit, which leads
to the surprising conclusion that for sufficiently large input sizes, the
difference between the QFT and the logarithmic-depth AQFT is negligible.  In
effect, the AQFT can be used as an direct replacement for the QFT, yielding
improvements in any application which does not require exact quantum
computation.

\end{abstract}

\section{Introduction}

The quantum Fourier transform is a fundamental component of many of the
significant quantum algorithms known to date.  
The main concern with the efficiency of the QFT algorithm is its use of 
controlled phase-shift gates which involve very small phases.  As the
number of qubits in the input, $n$, increases, the QFT algorithm requires
exponentially smaller phase shifts, which may be increasingly difficult or
even infeasible to implement physically.

Fortunately, for applications using the QFT which do not require an exact
answer, the idea of an approximate QFT was suggested\cite{Copp94}, in which
the phase shift gates requiring the most precision are omitted, since these 
are also the phase shift gates which have the least effect on the output.  
There is clearly a trade-off between the accuracy of the AQFT and its
circuit complexity.  Barenco, Ekert, Suominen and T\"{o}rm\"{a} established a
lower bound for the probability that the AQFT returns the correct answer, and
concluded that the accuracy of the QFT can be achieved by $O(n^3/m^3)$
iterations of the AQFT, where $n$ is the size of the input register, and $m$
is the depth of the AQFT circuit.  Usually, we have $m=O(\log n)$.

However, using a more thorough analysis of the AQFT circuit, we derive a much
better lower bound on the accuracy of the AQFT which allows us to conclude
that the accuracy of the QFT can be achieved by just $1+o(1)$ iterations of the
AQFT.

\section{The Quantum Fourier Transform}

The quantum Fourier transform is a quantum circuit which performs a discrete
Fourier transform on the complex-valued vector of $2^n$ probability amplitudes
associated with an $n$-qubit quantum system.  Specifically, given an
$n$-qubit state as a superposition of basis states $\ket{0}$, $\ket{1}$,
$\ldots$, $\ket{2^n-1}$, the QFT maps each basis state $\ket{j}$ to
$$QFT(\ket{j})=\frac{1}{\sqrt{2^n}}\sum_{k=0}^{2^n-1}e^{2\pi ijk/2^n}\ket{k}.$$

\begin{figure}
  \centerline{\hbox{\setlength{\unitlength}{3158sp}%
\begingroup\makeatletter\ifx\SetFigFont\undefined%
\gdef\SetFigFont#1#2#3#4#5{%
  \reset@font\fontsize{#1}{#2pt}%
  \fontfamily{#3}\fontseries{#4}\fontshape{#5}%
  \selectfont}%
\fi\endgroup%
\begin{picture}(8112,2724)(151,-1873)
\put(6376,-61){\makebox(0,0)[lb]{\smash{\SetFigFont{10}{12.0}{\rmdefault}{\mddefault}{\updefault}{\color[rgb]{0,0,0}$\left|0\right>+e^{2\pi i(0.x_2\cdots x_n)}\left|1\right>$}%
}}}
\put(751,539){\makebox(0,0)[lb]{\smash{\SetFigFont{10}{12.0}{\rmdefault}{\mddefault}{\updefault}{\color[rgb]{0,0,0}$H$}%
}}}
\thinlines
{\color[rgb]{0,0,0}\put(1201,389){\framebox(450,450){}}
}%
\put(1276,539){\makebox(0,0)[lb]{\smash{\SetFigFont{10}{12.0}{\rmdefault}{\mddefault}{\updefault}{\color[rgb]{0,0,0}$R_2$}%
}}}
{\color[rgb]{0,0,0}\put(1876,389){\framebox(450,450){}}
}%
\put(1876,539){\makebox(0,0)[lb]{\smash{\SetFigFont{10}{12.0}{\rmdefault}{\mddefault}{\updefault}{\color[rgb]{0,0,0}$R_{n-1}$}%
}}}
{\color[rgb]{0,0,0}\put(2476,389){\framebox(450,450){}}
}%
\put(2551,539){\makebox(0,0)[lb]{\smash{\SetFigFont{10}{12.0}{\rmdefault}{\mddefault}{\updefault}{\color[rgb]{0,0,0}$R_n$}%
}}}
{\color[rgb]{0,0,0}\put(2926,-211){\framebox(450,450){}}
}%
\put(3076,-61){\makebox(0,0)[lb]{\smash{\SetFigFont{10}{12.0}{\rmdefault}{\mddefault}{\updefault}{\color[rgb]{0,0,0}$H$}%
}}}
{\color[rgb]{0,0,0}\put(3601,-211){\framebox(450,450){}}
}%
\put(3601,-61){\makebox(0,0)[lb]{\smash{\SetFigFont{10}{12.0}{\rmdefault}{\mddefault}{\updefault}{\color[rgb]{0,0,0}$R_{n-2}$}%
}}}
{\color[rgb]{0,0,0}\put(4201,-211){\framebox(450,450){}}
}%
\put(4201,-61){\makebox(0,0)[lb]{\smash{\SetFigFont{10}{12.0}{\rmdefault}{\mddefault}{\updefault}{\color[rgb]{0,0,0}$R_{n-1}$}%
}}}
{\color[rgb]{0,0,0}\put(4651,-1261){\framebox(450,450){}}
}%
\put(4801,-1111){\makebox(0,0)[lb]{\smash{\SetFigFont{10}{12.0}{\rmdefault}{\mddefault}{\updefault}{\color[rgb]{0,0,0}$H$}%
}}}
{\color[rgb]{0,0,0}\put(5251,-1261){\framebox(450,450){}}
}%
\put(5326,-1111){\makebox(0,0)[lb]{\smash{\SetFigFont{10}{12.0}{\rmdefault}{\mddefault}{\updefault}{\color[rgb]{0,0,0}$R_2$}%
}}}
{\color[rgb]{0,0,0}\put(5701,-1861){\framebox(450,450){}}
}%
\put(5851,-1711){\makebox(0,0)[lb]{\smash{\SetFigFont{10}{12.0}{\rmdefault}{\mddefault}{\updefault}{\color[rgb]{0,0,0}$H$}%
}}}
{\color[rgb]{0,0,0}\put(1426, 14){\circle*{150}}
}%
{\color[rgb]{0,0,0}\put(2101,-1036){\circle*{150}}
}%
{\color[rgb]{0,0,0}\put(2701,-1636){\circle*{150}}
}%
{\color[rgb]{0,0,0}\put(3826,-1036){\circle*{150}}
}%
{\color[rgb]{0,0,0}\put(4426,-1636){\circle*{150}}
}%
{\color[rgb]{0,0,0}\put(5476,-1636){\circle*{150}}
}%
{\color[rgb]{0,0,0}\put(451,614){\line( 1, 0){150}}
}%
{\color[rgb]{0,0,0}\put(5701,-1636){\line(-1, 0){5250}}
}%
{\color[rgb]{0,0,0}\put(1051,614){\line( 1, 0){150}}
}%
{\color[rgb]{0,0,0}\multiput(1651,614)(45.00000,0.00000){6}{\makebox(2.0833,14.5833){\SetFigFont{5}{6}{\rmdefault}{\mddefault}{\updefault}.}}
}%
{\color[rgb]{0,0,0}\put(2926, 14){\line(-1, 0){2475}}
}%
{\color[rgb]{0,0,0}\put(1426,389){\line( 0,-1){375}}
}%
{\color[rgb]{0,0,0}\put(2101,389){\line( 0,-1){1425}}
}%
{\color[rgb]{0,0,0}\put(2326,614){\line( 1, 0){150}}
}%
{\color[rgb]{0,0,0}\put(2926,614){\line( 1, 0){3375}}
}%
{\color[rgb]{0,0,0}\put(2701,389){\line( 0,-1){2025}}
}%
{\color[rgb]{0,0,0}\multiput(3376, 14)(45.00000,0.00000){6}{\makebox(2.0833,14.5833){\SetFigFont{5}{6}{\rmdefault}{\mddefault}{\updefault}.}}
}%
{\color[rgb]{0,0,0}\put(4051, 14){\line( 1, 0){150}}
}%
{\color[rgb]{0,0,0}\put(3826,-211){\line( 0,-1){825}}
}%
{\color[rgb]{0,0,0}\put(4426,-211){\line( 0,-1){1425}}
}%
{\color[rgb]{0,0,0}\put(4651, 14){\line( 1, 0){1650}}
}%
{\color[rgb]{0,0,0}\put(4651,-1036){\line(-1, 0){3975}}
}%
{\color[rgb]{0,0,0}\put(5101,-1036){\line( 1, 0){150}}
}%
{\color[rgb]{0,0,0}\put(5476,-1261){\line( 0,-1){375}}
}%
{\color[rgb]{0,0,0}\put(5701,-1036){\line( 1, 0){600}}
}%
{\color[rgb]{0,0,0}\put(6151,-1636){\line( 1, 0){150}}
}%
{\color[rgb]{0,0,0}\multiput(6751,-211)(0.00000,-46.15385){14}{\makebox(2.0833,14.5833){\SetFigFont{5}{6}{\rmdefault}{\mddefault}{\updefault}.}}
}%
{\color[rgb]{0,0,0}\multiput(301,-211)(0.00000,-46.15385){14}{\makebox(2.0833,14.5833){\SetFigFont{5}{6}{\rmdefault}{\mddefault}{\updefault}.}}
}%
{\color[rgb]{1,1,1}\put(8251,-211){\line( 0,-1){600}}
}%
\put(151,539){\makebox(0,0)[lb]{\smash{\SetFigFont{10}{12.0}{\rmdefault}{\mddefault}{\updefault}{\color[rgb]{0,0,0}$\left|x_1\right>$}%
}}}
\put(151,-61){\makebox(0,0)[lb]{\smash{\SetFigFont{10}{12.0}{\rmdefault}{\mddefault}{\updefault}{\color[rgb]{0,0,0}$\left|x_2\right>$}%
}}}
\put(151,-1111){\makebox(0,0)[lb]{\smash{\SetFigFont{10}{12.0}{\rmdefault}{\mddefault}{\updefault}{\color[rgb]{0,0,0}$\left|x_{n-1}\right>$}%
}}}
\put(151,-1711){\makebox(0,0)[lb]{\smash{\SetFigFont{10}{12.0}{\rmdefault}{\mddefault}{\updefault}{\color[rgb]{0,0,0}$\left|x_n\right>$}%
}}}
\put(6376,-1111){\makebox(0,0)[lb]{\smash{\SetFigFont{10}{12.0}{\rmdefault}{\mddefault}{\updefault}{\color[rgb]{0,0,0}$\left|0\right>+e^{2\pi i(0.x_{n-1}x_n)}\left|1\right>$}%
}}}
\put(6376,-1711){\makebox(0,0)[lb]{\smash{\SetFigFont{10}{12.0}{\rmdefault}{\mddefault}{\updefault}{\color[rgb]{0,0,0}$\left|0\right>+e^{2\pi i(0.x_n)}\left|1\right>$}%
}}}
\put(6376,539){\makebox(0,0)[lb]{\smash{\SetFigFont{10}{12.0}{\rmdefault}{\mddefault}{\updefault}{\color[rgb]{0,0,0}$\left|0\right>+e^{2\pi i(0.x_1x_2\cdots x_n)}\left|1\right>$}%
}}}
{\color[rgb]{0,0,0}\put(601,389){\framebox(450,450){}}
}%
\end{picture}}}
  \caption{A quantum circuit that performs a quantum Fourier transform}
  \label{qftfig1}
\end{figure}
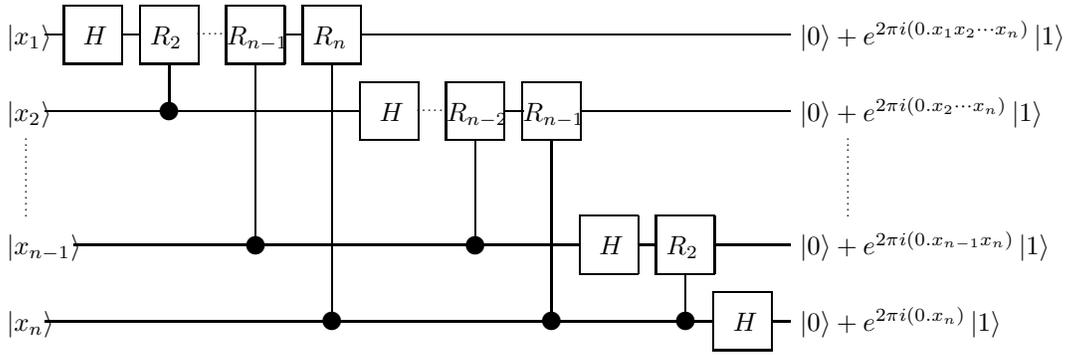

Figure \ref{qftfig1} gives a quantum circuit which performs a QFT.  The input
register contains an $n$-qubit basis state
$\ket{x}\equiv\ket{x_1x_2\cdots x_n}$.
The gates labelled $H$ are Hadamard gates,
and the gates labelled $R_m$ represent a series of one-qubit
phase rotation gates.  For each integer $m\geq 2$, the gate $R_m$ shifts the
phase of the $\ket{1}$ component of the input qubit by a factor of
$e^{2\pi i/2^m}$.  Note that the output state of the QFT can be factored into
a tensor product as indicated in the diagram.

In practice, the QFT is usually used in reverse, where the states
$$\frac{1}{\sqrt{2}}(\ket{0}+e^{2\pi i(2^{p-1}\phi)}\ket{1})$$
are prepared for $p=1,2,\ldots,n$ by some other means, and used as input
into the inverse QFT circuit in order to obtain an estimate of the value of
$\phi$ from the state $\ket{x}=\ket{x_1x_2\cdots x_n}$.  Specifically, $\phi$
is estimated by $\hat{\phi}=(0.x_1x_2\ldots x_n)$, where the parentheses
denote a binary (base-2) fraction.  The estimate $\hat{\phi}$ is the nearest
multiple of $2^{-n}$ to $\phi$ with probability $\frac{4}{\pi^2}$, and one of
the two nearest multiples with probability $\frac{8}{\pi^2}$.

As for the generation of the required input states, since it is not important
how they are generated, we may assume that they are generated by using the
state $\frac{1}{\sqrt{2}}(\ket{0}+\ket{1})$ as the control qubit for a
controlled-$U^{2^{p-1}}$ gate, where $U$ is a quantum gate with $\phi$ as an
eigenvalue.

\begin{figure}
  \centerline{\hbox{\setlength{\unitlength}{3947sp}%
\begingroup\makeatletter\ifx\SetFigFont\undefined%
\gdef\SetFigFont#1#2#3#4#5{%
  \reset@font\fontsize{#1}{#2pt}%
  \fontfamily{#3}\fontseries{#4}\fontshape{#5}%
  \selectfont}%
\fi\endgroup%
\begin{picture}(3150,1074)(1,-223)
\put(  1,-61){\makebox(0,0)[lb]{\smash{\SetFigFont{12}{14.4}{\rmdefault}{\mddefault}{\updefault}{\color[rgb]{0,0,0}$\left|u\right>$}%
}}}
\put(751,539){\makebox(0,0)[lb]{\smash{\SetFigFont{12}{14.4}{\rmdefault}{\mddefault}{\updefault}{\color[rgb]{0,0,0}$H$}%
}}}
\thinlines
{\color[rgb]{0,0,0}\put(2476,389){\framebox(450,450){}}
}%
\put(2626,539){\makebox(0,0)[lb]{\smash{\SetFigFont{12}{14.4}{\rmdefault}{\mddefault}{\updefault}{\color[rgb]{0,0,0}$H$}%
}}}
{\color[rgb]{0,0,0}\put(1426,614){\circle*{150}}
}%
{\color[rgb]{0,0,0}\put(451,614){\line( 1, 0){150}}
}%
{\color[rgb]{0,0,0}\put(1051,614){\line( 1, 0){750}}
}%
{\color[rgb]{0,0,0}\put(1201, 14){\line(-1, 0){750}}
}%
{\color[rgb]{0,0,0}\put(1201, 89){\line(-1, 0){750}}
}%
{\color[rgb]{0,0,0}\put(1201,-61){\line(-1, 0){750}}
}%
{\color[rgb]{0,0,0}\put(1426,239){\line( 0, 1){375}}
}%
{\color[rgb]{0,0,0}\put(1651, 89){\line( 1, 0){150}}
}%
{\color[rgb]{0,0,0}\put(1651, 14){\line( 1, 0){150}}
}%
{\color[rgb]{0,0,0}\put(1651,-61){\line( 1, 0){150}}
}%
{\color[rgb]{0,0,0}\put(1801,389){\framebox(525,450){}}
}%
{\color[rgb]{0,0,0}\put(2326,614){\line( 1, 0){150}}
}%
{\color[rgb]{0,0,0}\put(2926,614){\line( 1, 0){150}}
}%
{\color[rgb]{0,0,0}\put(1201,-211){\framebox(450,450){}}
}%
\put(  1,539){\makebox(0,0)[lb]{\smash{\SetFigFont{12}{14.4}{\rmdefault}{\mddefault}{\updefault}{\color[rgb]{0,0,0}$\left|0\right>$}%
}}}
\put(1201,-61){\makebox(0,0)[lb]{\smash{\SetFigFont{12}{14.4}{\rmdefault}{\mddefault}{\updefault}{\color[rgb]{0,0,0}$U^{2^{p-1}}$}%
}}}
\put(3151,539){\makebox(0,0)[lb]{\smash{\SetFigFont{12}{14.4}{\rmdefault}{\mddefault}{\updefault}{\color[rgb]{0,0,0}$\left|x_p\right>$}%
}}}
\put(1876,539){\makebox(0,0)[lb]{\smash{\SetFigFont{12}{14.4}{\rmdefault}{\mddefault}{\updefault}{\color[rgb]{0,0,0}$R_{-\chi_p}$}%
}}}
\put(1876,-61){\makebox(0,0)[lb]{\smash{\SetFigFont{12}{14.4}{\rmdefault}{\mddefault}{\updefault}{\color[rgb]{0,0,0}$\left|u\right>$}%
}}}
{\color[rgb]{0,0,0}\put(601,389){\framebox(450,450){}}
}%
\end{picture}}}
  \caption{An individual trial}
  \label{qftfig8b}
\end{figure}
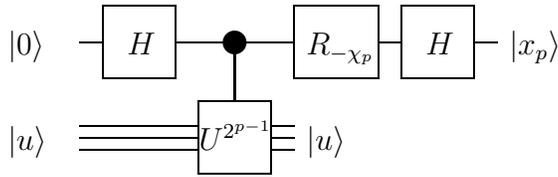

Griffiths and Niu observed that it is not necessary for the entire inverse QFT
to be performed at once\cite{GN96}.  Instead, since each output qubit,
$\ket{x_p}$ is controlled only by higher-indexed qubits, and the last qubit
$\ket{x_n}$ can be obtained independently of all other qubits, they suggested
a ``semiclassical'' version of the inverse QFT circuit in which each output
qubit is obtained one at time, starting with $\ket{x_n}$.  Proceeding one
qubit at a time, the results from measuring the previous qubits are used to
classically determine whether a particular phase rotation gate will be
applied to the current qubit.  This process does not affect the final answer,
but has the advantage of replacing the controlled-rotation gates in the
inverse QFT with regular rotation gates.  Thus, we can analyze the entire
inverse QFT circuit simply by looking at each individual ``trial''.

In the circuit in Figure \ref{qftfig8b}, the controlled rotation gates are
replaced with a rotation by a phase of $-\chi_p$, which is determined by the
results of the previous trials, from $\ket{x_n}$ to $\ket{x_{p+1}}$.
Specifically, $\chi_p=(0.0x_{p+1}\ldots x_n)$.

\section{Approximate QFT Circuits}

In Coppersmith's AQFT circuits\cite{Copp94}, instead of performing all phase
rotations $R_p$ that are used in the QFT algorithm, we set a lower limit on
the amount of phase shifted by any particular gate and ignore any phase
rotation gates that do not shift by at least that amount.  In other words,
given a positive integer threshold, $m$, we will define the approximate QFT
circuit AQFT$_m$ to be the circuit formed from the QFT, except that phase
rotation gates $R_p$ are ignored whenever we have $m>p$.  Note that AQFT$_n$
is simply the regular QFT circuit.  As before, the inverse AQFT can be divided
into individual trials.

Suppose we are given a phase $\phi$, and we are to use the AQFT$_m$ circuit to
find an approximation $\hat{\phi}=(0.x_1x_2\ldots x_n)$.  The individual trial
for bit $x_p$ starts with the input state
$\ket{0}+e^{2\pi i(2^{p-1}\phi)}\ket{1}$, and performs a phase rotation of
$e^{-2\pi i\chi_p}$ on the probability amplitude of $\ket{1}$.  Here, $\chi_p$
corresponds to the phase rotation amount performed by the regular QFT circuit,
$(0.0x_{i+1}x_{i+2}\ldots x_n)$,
but without the effect of any gates which are smaller than the given threshold
$m$.  Therefore, we have $\chi_p=(0.0x_{p+1}\ldots x_{p+m-1})$ when
$p\leq n-m$, and $\chi_p=(0.0x_{p+1}\ldots x_n)$ otherwise.

We will now compute the probability that $\hat{\phi}$ is the nearest integer
multiple of $1/2^n$ to $\phi$, i.e., $|\phi-\hat{\phi}|\leq 2^{-n-1}$.
First, using the double angle formula and induction, we can easily establish
the fact that for any integer $n\geq 1$, we have
$$\prod_{p=1}^{n}\cos^2\frac{\theta}{2^p}=%
\left(\frac{\sin \theta}{2^n\sin\frac{\theta}{2^n}}\right)^2.$$
Now observe that, in general, applying a Hadamard gate to the state
$\ket{0}+e^{2\pi i((0.x_p)+\delta_p)}\ket{1}$ and measuring yields $\ket{x_p}$
with probability $\cos^2\left(\pi\delta_p\right)$.  However, for the
individual AQFT$_m$ trial for bit $x_p$, we know what $\delta_p$ has to be.
We have two cases.

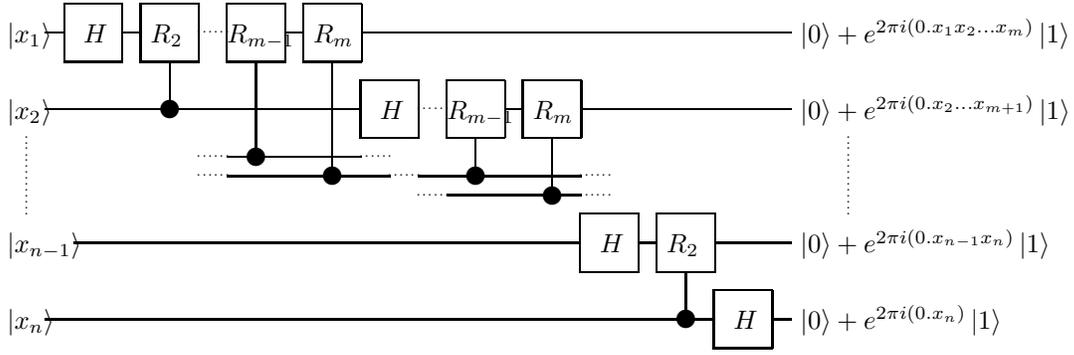
\begin{figure}
  \centerline{\setlength{\unitlength}{3158sp}%
\begingroup\makeatletter\ifx\SetFigFont\undefined%
\gdef\SetFigFont#1#2#3#4#5{%
  \reset@font\fontsize{#1}{#2pt}%
  \fontfamily{#3}\fontseries{#4}\fontshape{#5}%
  \selectfont}%
\fi\endgroup%
\begin{picture}(8262,2724)(151,-1873)
\put(6376,539){\makebox(0,0)[lb]{\smash{\SetFigFont{10}{12.0}{\rmdefault}{\mddefault}{\updefault}{\color[rgb]{0,0,0}$\left|0\right>+e^{2\pi i(0.x_1x_2\ldots x_m)}\left|1\right>$}%
}}}
\put(751,539){\makebox(0,0)[lb]{\smash{\SetFigFont{10}{12.0}{\rmdefault}{\mddefault}{\updefault}{\color[rgb]{0,0,0}$H$}%
}}}
\thinlines
{\color[rgb]{0,0,0}\put(1201,389){\framebox(450,450){}}
}%
\put(1276,539){\makebox(0,0)[lb]{\smash{\SetFigFont{10}{12.0}{\rmdefault}{\mddefault}{\updefault}{\color[rgb]{0,0,0}$R_2$}%
}}}
{\color[rgb]{0,0,0}\put(2926,-211){\framebox(450,450){}}
}%
\put(3076,-61){\makebox(0,0)[lb]{\smash{\SetFigFont{10}{12.0}{\rmdefault}{\mddefault}{\updefault}{\color[rgb]{0,0,0}$H$}%
}}}
{\color[rgb]{0,0,0}\put(4651,-1261){\framebox(450,450){}}
}%
\put(4801,-1111){\makebox(0,0)[lb]{\smash{\SetFigFont{10}{12.0}{\rmdefault}{\mddefault}{\updefault}{\color[rgb]{0,0,0}$H$}%
}}}
{\color[rgb]{0,0,0}\put(5251,-1261){\framebox(450,450){}}
}%
\put(5326,-1111){\makebox(0,0)[lb]{\smash{\SetFigFont{10}{12.0}{\rmdefault}{\mddefault}{\updefault}{\color[rgb]{0,0,0}$R_2$}%
}}}
{\color[rgb]{0,0,0}\put(5701,-1861){\framebox(450,450){}}
}%
\put(5851,-1711){\makebox(0,0)[lb]{\smash{\SetFigFont{10}{12.0}{\rmdefault}{\mddefault}{\updefault}{\color[rgb]{0,0,0}$H$}%
}}}
{\color[rgb]{0,0,0}\put(1876,389){\framebox(450,450){}}
}%
{\color[rgb]{0,0,0}\put(1426, 14){\circle*{150}}
}%
{\color[rgb]{0,0,0}\put(5476,-1636){\circle*{150}}
}%
{\color[rgb]{0,0,0}\put(2101,-361){\circle*{150}}
}%
{\color[rgb]{0,0,0}\put(2701,-511){\circle*{150}}
}%
{\color[rgb]{0,0,0}\put(3826,-511){\circle*{150}}
}%
{\color[rgb]{0,0,0}\put(4426,-661){\circle*{150}}
}%
{\color[rgb]{0,0,0}\put(451,614){\line( 1, 0){150}}
}%
{\color[rgb]{0,0,0}\put(5701,-1636){\line(-1, 0){5250}}
}%
{\color[rgb]{0,0,0}\put(1051,614){\line( 1, 0){150}}
}%
{\color[rgb]{0,0,0}\put(2926, 14){\line(-1, 0){2475}}
}%
{\color[rgb]{0,0,0}\put(1426,389){\line( 0,-1){375}}
}%
{\color[rgb]{0,0,0}\put(2101,389){\line( 0,-1){750}}
}%
{\color[rgb]{0,0,0}\put(2326,614){\line( 1, 0){150}}
}%
{\color[rgb]{0,0,0}\put(2926,614){\line( 1, 0){3375}}
}%
{\color[rgb]{0,0,0}\put(2701,389){\line( 0,-1){900}}
}%
{\color[rgb]{0,0,0}\multiput(3376, 14)(45.00000,0.00000){6}{\makebox(2.0833,14.5833){\SetFigFont{5}{6}{\rmdefault}{\mddefault}{\updefault}.}}
}%
{\color[rgb]{0,0,0}\put(4051, 14){\line( 1, 0){150}}
}%
{\color[rgb]{0,0,0}\put(3826,-211){\line( 0,-1){300}}
}%
{\color[rgb]{0,0,0}\put(4426,-211){\line( 0,-1){450}}
}%
{\color[rgb]{0,0,0}\put(4651, 14){\line( 1, 0){1650}}
}%
{\color[rgb]{0,0,0}\put(4651,-1036){\line(-1, 0){3975}}
}%
{\color[rgb]{0,0,0}\put(5101,-1036){\line( 1, 0){150}}
}%
{\color[rgb]{0,0,0}\put(5476,-1261){\line( 0,-1){375}}
}%
{\color[rgb]{0,0,0}\put(5701,-1036){\line( 1, 0){600}}
}%
{\color[rgb]{0,0,0}\put(6151,-1636){\line( 1, 0){150}}
}%
{\color[rgb]{0,0,0}\multiput(6751,-211)(0.00000,-46.15385){14}{\makebox(2.0833,14.5833){\SetFigFont{5}{6}{\rmdefault}{\mddefault}{\updefault}.}}
}%
{\color[rgb]{0,0,0}\put(1876,-361){\line( 1, 0){1050}}
}%
{\color[rgb]{0,0,0}\put(1876,-511){\line( 1, 0){1275}}
}%
{\color[rgb]{0,0,0}\put(3376,-511){\line( 1, 0){1275}}
}%
{\color[rgb]{0,0,0}\put(3601,-661){\line( 1, 0){1050}}
}%
{\color[rgb]{0,0,0}\multiput(3151,-511)(45.00000,0.00000){6}{\makebox(2.0833,14.5833){\SetFigFont{5}{6}{\rmdefault}{\mddefault}{\updefault}.}}
}%
{\color[rgb]{0,0,0}\multiput(1651,-361)(45.00000,0.00000){6}{\makebox(2.0833,14.5833){\SetFigFont{5}{6}{\rmdefault}{\mddefault}{\updefault}.}}
}%
{\color[rgb]{0,0,0}\multiput(1651,-511)(45.00000,0.00000){6}{\makebox(2.0833,14.5833){\SetFigFont{5}{6}{\rmdefault}{\mddefault}{\updefault}.}}
}%
{\color[rgb]{0,0,0}\multiput(2926,-361)(45.00000,0.00000){6}{\makebox(2.0833,14.5833){\SetFigFont{5}{6}{\rmdefault}{\mddefault}{\updefault}.}}
}%
{\color[rgb]{0,0,0}\multiput(4651,-511)(45.00000,0.00000){6}{\makebox(2.0833,14.5833){\SetFigFont{5}{6}{\rmdefault}{\mddefault}{\updefault}.}}
}%
{\color[rgb]{0,0,0}\multiput(4651,-661)(45.00000,0.00000){6}{\makebox(2.0833,14.5833){\SetFigFont{5}{6}{\rmdefault}{\mddefault}{\updefault}.}}
}%
{\color[rgb]{0,0,0}\multiput(3376,-661)(45.00000,0.00000){6}{\makebox(2.0833,14.5833){\SetFigFont{5}{6}{\rmdefault}{\mddefault}{\updefault}.}}
}%
{\color[rgb]{0,0,0}\multiput(1651,614)(45.00000,0.00000){6}{\makebox(2.0833,14.5833){\SetFigFont{5}{6}{\rmdefault}{\mddefault}{\updefault}.}}
}%
{\color[rgb]{0,0,0}\put(2476,389){\framebox(450,450){}}
}%
{\color[rgb]{0,0,0}\put(3601,-211){\framebox(450,450){}}
}%
{\color[rgb]{0,0,0}\put(4201,-211){\framebox(450,450){}}
}%
{\color[rgb]{0,0,0}\multiput(301,-211)(0.00000,-46.15385){14}{\makebox(2.0833,14.5833){\SetFigFont{5}{6}{\rmdefault}{\mddefault}{\updefault}.}}
}%
{\color[rgb]{1,1,1}\put(8401,-61){\line( 0,-1){1200}}
}%
\put(1876,539){\makebox(0,0)[lb]{\smash{\SetFigFont{10}{12.0}{\rmdefault}{\mddefault}{\updefault}{\color[rgb]{0,0,0}$R_{m-1}$}%
}}}
\put(2551,539){\makebox(0,0)[lb]{\smash{\SetFigFont{10}{12.0}{\rmdefault}{\mddefault}{\updefault}{\color[rgb]{0,0,0}$R_m$}%
}}}
\put(3601,-61){\makebox(0,0)[lb]{\smash{\SetFigFont{10}{12.0}{\rmdefault}{\mddefault}{\updefault}{\color[rgb]{0,0,0}$R_{m-1}$}%
}}}
\put(4276,-61){\makebox(0,0)[lb]{\smash{\SetFigFont{10}{12.0}{\rmdefault}{\mddefault}{\updefault}{\color[rgb]{0,0,0}$R_m$}%
}}}
\put(151,539){\makebox(0,0)[lb]{\smash{\SetFigFont{10}{12.0}{\rmdefault}{\mddefault}{\updefault}{\color[rgb]{0,0,0}$\left|x_1\right>$}%
}}}
\put(151,-61){\makebox(0,0)[lb]{\smash{\SetFigFont{10}{12.0}{\rmdefault}{\mddefault}{\updefault}{\color[rgb]{0,0,0}$\left|x_2\right>$}%
}}}
\put(151,-1111){\makebox(0,0)[lb]{\smash{\SetFigFont{10}{12.0}{\rmdefault}{\mddefault}{\updefault}{\color[rgb]{0,0,0}$\left|x_{n-1}\right>$}%
}}}
\put(151,-1711){\makebox(0,0)[lb]{\smash{\SetFigFont{10}{12.0}{\rmdefault}{\mddefault}{\updefault}{\color[rgb]{0,0,0}$\left|x_n\right>$}%
}}}
\put(6376,-1711){\makebox(0,0)[lb]{\smash{\SetFigFont{10}{12.0}{\rmdefault}{\mddefault}{\updefault}{\color[rgb]{0,0,0}$\left|0\right>+e^{2\pi i(0.x_n)}\left|1\right>$}%
}}}
\put(6376,-1111){\makebox(0,0)[lb]{\smash{\SetFigFont{10}{12.0}{\rmdefault}{\mddefault}{\updefault}{\color[rgb]{0,0,0}$\left|0\right>+e^{2\pi i(0.x_{n-1}x_n)}\left|1\right>$}%
}}}
\put(6376,-61){\makebox(0,0)[lb]{\smash{\SetFigFont{10}{12.0}{\rmdefault}{\mddefault}{\updefault}{\color[rgb]{0,0,0}$\left|0\right>+e^{2\pi i(0.x_2\ldots x_{m+1})}\left|1\right>$}%
}}}
{\color[rgb]{0,0,0}\put(601,389){\framebox(450,450){}}
}%
\end{picture}}
  \caption{A quantum circuit implementing the AQFT}
  \label{qftfig5}
\end{figure}

In the first case, where $p+m-1<n$ and $\chi_p=(0.0x_{p+1}\ldots x_{p+m-1})$,
$\delta_p$ contains the phase rotation amount that was not
applied by the AQFT$_m$ trial that would have been applied by the
corresponding QFT trial, and the difference between $\phi$ and its nearest
integer estimate, amplified by a factor of $2^p$ by the AQFT$_m$ circuit.  In
other words, we have
$$\delta_p=2^{-m}(0.x_{p+m}x_{p+m+1}\ldots x_n)+2^{p-1}\delta,$$
where $\delta=\phi-\hat{\phi}$.  Also, if $\hat{\phi}$ is the nearest estimate
of $\phi$, we have $$-2^{p-n-1}\leq2^{p-1}\delta\leq 2^{p-n-1}.$$
Now, since $(0.x_{p+m}\ldots x_n)$
reaches its minimum value when all the bits are $0$, and its maximum value
when all its bits are $1$, we have
$$0\leq 2^{-m}(0.x_{p+m}\ldots x_n)\leq
2^{-m}(1-2^{p+m-n-1})=2^{-m}-2^{p-n-1}.$$ 
Adding the two inequalities together, we obtain
$-2^{-m}<-2^{-n-1}\leq\delta_p\leq2^{-m}$.  This means that final measurement
will yield $\ket{x_p}$ with a probability of at least $\cos^2(\pi2^{-m})$ in
this case.

In the second case, where $p\geq n-m+1$, we simply have
$\delta_p=2^{p-1}\delta$.
Since $|\delta|\leq 2^{-n-1}$, we have an error probability lower bound of
$\cos^2(\frac{\pi}{2}2^{p-n-1})$.
The probability $P$ of getting the
correct output with the AQFT$_m$ circuit is simply the product of each
individual bit trial being correct.  Putting these lower bounds together, we
can give a lower bound for $P$:
\begin{eqnarray*}
P&\geq&\prod_{p=n-m+1}^n\cos^2\left(\frac{\pi}{2}2^{p-n-1}\right)
\prod_{p=1}^{n-m}\cos^2(\pi 2^{-m})\\
&=&\left(\frac{\sin\frac{\pi}{2}}{2^m\sin\frac{\pi/2}{2^m}}\right)^2
\left(\cos^2(\pi 2^{-m})\right)^{n-m}\\
&\geq&\left(\frac{1}{\pi/2}\right)^2
\left(\cos^2(\pi 2^{-m})\right)^{n-m}\\
&=&\frac{4}{\pi^2}\left(\cos^2(\pi 2^{-m})\right)^{n-m}.
\end{eqnarray*}

It is clear that if $m$ grows too slowly in comparison to $n$, then the
expression will approach $0$ asymptotically as $n$ approaches $\infty$, making
this bound useless.  If we take $m\geq\log_2n+2$ so that $2^m\geq4n$, we would
have
$$P\geq\frac{4}{\pi^2}\left(\cos^2(\pi 2^{-m})\right)^{n-m}
\geq\frac{4}{\pi^2}\left(\cos^2\left(\frac{\pi}{4n}\right)\right)^n,$$
giving us
$$\lim_{n\rightarrow\infty}P\geq\lim_{n\rightarrow\infty}
\frac{4}{\pi^2}\left(\cos^2\left(\frac{\pi}{4n}\right)\right)^n
=\frac{4}{\pi^2}.$$
Since Cleve and Watrous\cite{CW} establish a lower bound of $\Omega(\log n)$
on the depth of any AQFT circuit with a constant error bound, requiring 
$m\geq\log_2n+2$ is a reasonable restriction on a logarithmic-depth AQFT
circuit.

Asymptotically, for $m\geq\log_2n+2$, the lower bound for the
accuracy of the AQFT$_m$ circuit approaches the lower bound for the accuracy
of the full QFT, meaning that for sufficiently large $n$, the QFT may be
substituted by the AQFT$_m$ circuit with only a negligible effect.
This is a significant improvement, since the full QFT on an $n$-qubit
register requires $O(n^2)$ gates to implement, while the AQFT$_m$ circuit uses
only $O(nm)=O(n\log_2n)$ gates to achieve nearly the same level of accuracy.

Finally, we will also establish a lower bound for $P$ when we have
$m\geq\log_2n+2$ for some fixed value of $n$, since an asymptotic bound gives
no information about the value of $P$ for practical values of $n$.  We have
\begin{eqnarray*}
P&\geq&\frac{4}{\pi^2}\left(\cos^2\left(\frac{\pi}{4n}\right)\right)^{n-m}\\
&=&\frac{4}{\pi^2}\left(1-\sin^2\left(\frac{\pi}{4n}\right)\right)^{n-m}\\
&\geq&\frac{4}{\pi^2}\left(1-(n-m)\sin^2\left(\frac{\pi}{4n}\right)\right),
\end{eqnarray*}
using the Bernoulli inequality, which states that
$(1+t)^c\geq1+ct$ whenever $t>-1$ and $c\geq 0$.
Since $x\geq\sin x$ when $x\geq 0$, we have
\begin{eqnarray*}
P&\geq&\frac{4}{\pi^2}\left(1-(n-m)\left(\frac{\pi}{4n}\right)^2\right)\\
&\geq&\frac{4}{\pi^2}-\frac{1}{4n}.
\end{eqnarray*}
This indicates that even for smaller values of $n$, the probability of the
AQFT$_m$ circuit returning the best estimate of $\phi$ is high enough to make
it a practical alternative to using the full QFT circuit.

Note that since we stipulated that $m\geq\log_2n+2$, this result is only
meaningful if $n\geq 4$.  Otherwise, the logarithmic-depth AQFT$_m$ circuit is
simply the full QFT, with lower bound $P\geq\frac{4}{\pi^2}$.
We can now establish the fixed bound
$$P\geq\frac{4}{\pi^2}-\frac{1}{16}$$
for AQFT$_m$ circuits where $m\geq\log_2n+2$.

\section{Conclusion}

The fixed bound of $P\geq\frac{4}{\pi^2}-\frac{1}{16}$ derived here is much
stronger than the bound of
$$P\geq\frac{8}{\pi^2}\sin^2\left(\frac{\pi m}{4n}\right)$$
previously established by Barenco {\em et al.} in \cite{BEST96}.  They
conclude that $O(n^3/m^3)$ iterations are necessary which means that the
AQFT$_m$ yields an advantage only when the QFT itself is affected in the
presence of decoherence, as the AQFT$_m$ would be less susceptible.

The new analysis presented in this paper establishes the effectiveness of the
AQFT$_m$ as a direct replacement for the QFT for a suitable size of input
register.  Furthermore, since we have
$$\lim_{n\rightarrow\infty}
\frac{8}{\pi^2}\sin^2\left(\frac{\pi m}{4n}\right)=0$$
when $m$ is fixed at $\log_2n+2$, it was not previously known that the
AQFT$_m$ circuit would perform so well for large $n$.
This is a significant result for the feasibility of any large-scale quantum
computation requiring the use of the QFT circuit.


\end{document}